\documentclass[prb,aps,twocolumn,amsmath,superscriptaddress,nofootnoteinbib]{revtex4}

\usepackage{graphicx}
\usepackage{epsfig}
\usepackage{gensymb}
\usepackage{physics}
\begin{document}

\title{Suppression of Decoherence tied to Electron-Phonon Coupling in Telecom-Compatible Quantum Dots: Low-threshold Reappearance Regime for Quantum State Inversion}

\author{A. Ramachandran}
\affiliation{Department of Physics and Atmospheric Science,
Dalhousie University, Halifax, Nova Scotia B3H 4R2 Canada}

\author{G. R. Wilbur}

\affiliation{Department of Physics and Atmospheric Science,
Dalhousie University, Halifax, Nova Scotia B3H 4R2 Canada}

\author{S. O'Neal\footnote{Present Address: Imec, Kissimmee, Florida, 34744, USA}}

\affiliation{The College of Optics and Photonics, University of Central Florida, Orlando, Florida 32816-2700, USA}

\author{D. G. Deppe\footnote{Present Address: SdPhotonics, Richardson, Texas 75081, USA}}

\affiliation{The College of Optics and Photonics, University of Central Florida, Orlando, Florida 32816-2700, USA}

\author{K. C. Hall\footnote{Corresponding Author: Kimberley.Hall@dal.ca}}

\affiliation{Department of Physics and Atmospheric Science,
Dalhousie University, Halifax, Nova Scotia B3H 4R2 Canada}




\begin{abstract}
We demonstrate suppression of dephasing tied to deformation potential coupling of confined electrons to longitunidal acoustic (LA) phonons in optical control experiments on large semiconductor quantum dots (QDs) with emission compatible with the low-dispersion telecommunications band at 1.3~$\mu$m.  By exploiting the sensitivity of the electron-phonon spectral density to the size and shape of the QD, we demonstrate a four-fold reduction in the threshold pulse area required to enter the decoupled regime for exciton inversion using adiabatic rapid passage (ARP).  Our calculations of the quantum state dynamics indicate that the symmetry of the QD wave function provides an additional means to engineer the electron-phonon interaction.  Our findings will support the development of solid-state quantum emitters in future distributed quantum networks using semiconductor QDs. 
\end{abstract}

\maketitle

A quantum emitter is a physical system that can be used to encode a quantum state via some internal degree of freedom (e.g. exciton, electron spin, valley) and is coupled to light via a dipolar transition that enables the conversion of that quantum state into the state of a photon and vice versa.  Such quantum emitters can be applied to quantum light sources for quantum cryptography or quantum imaging \cite{Senellart:2017,Huber:2018} and a collection of coupled quantum emitters can be used to realize a small quantum simulator or quantum memory node in a distributed quantum network \cite{Atature:2018}. Among solid-state quantum emitter systems, semiconductor QDs are particularly attractive due to their high radiative quantum efficiencies \cite{Atature:2018}, their strong optical coupling \cite{Eliseev:2000}, the ability to achieve inter-QD interactions \cite{Imamoglu:1999,Clark:2007,Kim:2011,Greilich:2011,Muller:2011}, and their tunable emission in the range of standard telecommunication wavelengths essential for long-distance quantum information transfer \cite{Senellart:2017,Park:2000,Haffouz:2018}.  The performance of QD-based single and entangled photon sources has continued to improve over the past decade, with the latter recently surpassing the commercial standard based on parametric down conversion \cite{Senellart:2017,Huber:2018}.  Progress towards the development of functional quantum networks using QDs has also occurred at a remarkable pace, including demonstrations of fast \cite{Mathew:2014} and arbitrary \cite{deVasconcellos:2010,Gamouras:2013,Mathew:2015,Widhalm:2018,Mathew:2011,Gamouras:2012} single qubit rotations, two-qubit gates \cite{Kim:2011,Muller:2011,Boyle:2008}, spin-photon entanglement \cite{DeGrave:2012,Gao:2012}, spin-spin entanglement \cite{Delteil:2016,Stockhill:2017}, and quantum state transfer \cite{He:2017}.   

\begin{figure}[!t]
	\includegraphics[width=0.99\columnwidth]{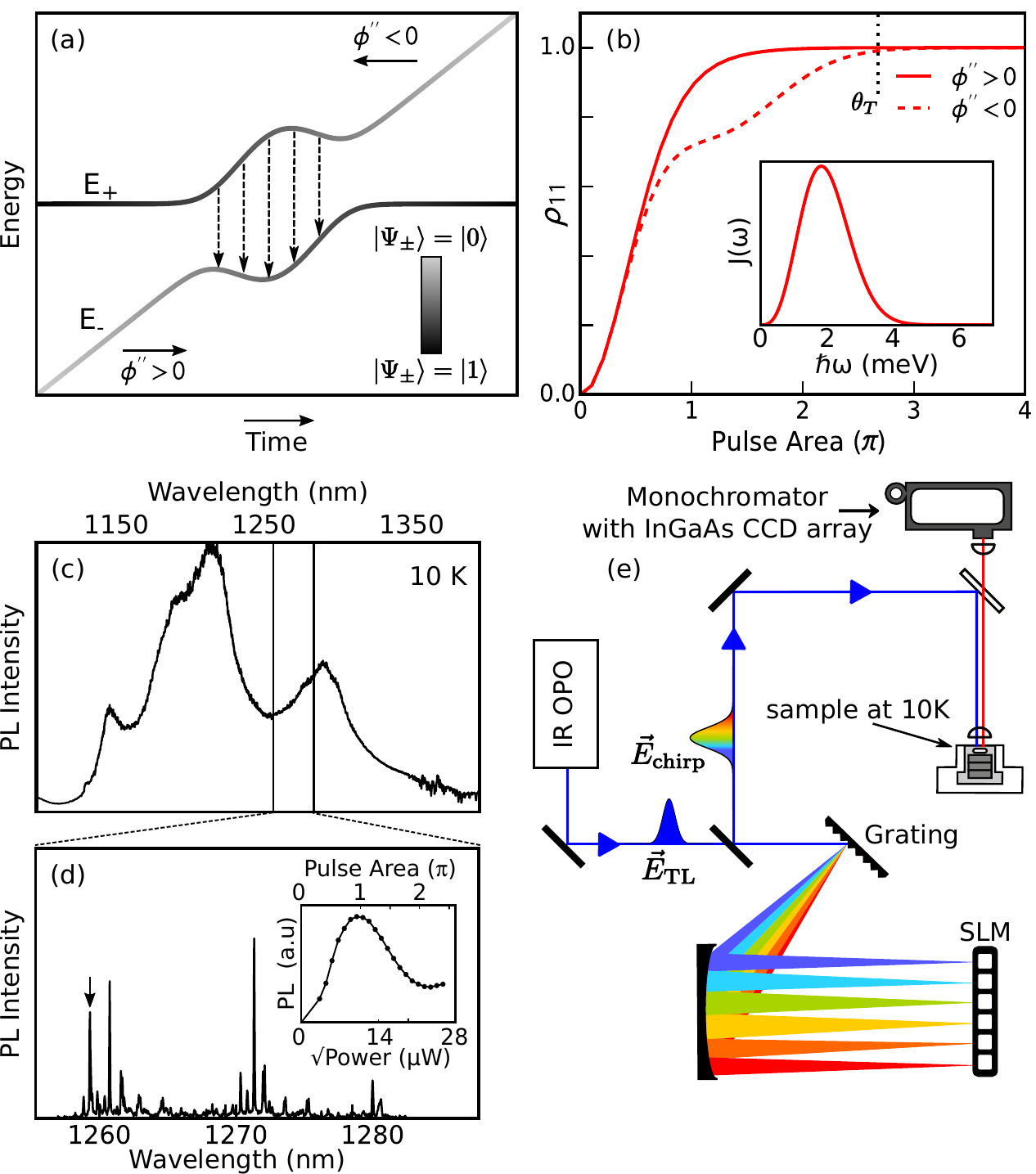}
    \caption{(a) The temporal evolution of the energies ($E_{\pm}$) of the dressed states ($\ket{\Psi_\pm}$) during ARP induced by a frequency-swept optical pulse. Dashed arrows indicate transitions between the dressed states caused by LA phonon emission.   (b) Calculated occupation of the GS exciton ($\ket{1}$) as a function of pulse area for positive (solid) and negative (dashed) pulse chirp corresponding to a spherical QD with $\omega_c$~=~1.49~eV ($d$~=~6.4~nm).  Inset: phonon spectral density.  (c) PL spectrum of the QD sample under pumping at 830~nm. (d) Micro-PL showing the GS emission from individual QDs. Inset: Rabi rotation measurement on the QD marked with the arrow.  The position of the first peak with respect to laser power is used to determine the pulse area for each QD. (e) Schematic diagram of the quantum control apparatus.       
}
    \label{fig:Figure1}
\end{figure}

The performance of the above technologies using QDs is limited by dephasing caused by coupling of the quantum-confined electrons with phonons in the solid-state environment. For optically-mediated quantum state control, the dominant decoherence channel is deformation potential coupling to longitudinal acoustic (LA) phonons, which reduces the fidelity of state manipulation through resonant phonon-induced transitions between the dressed states of the QD in the presence of the laser field \cite{LukerReview,Wigger:2018}.  These transitions lead to a strong damping of Rabi oscillations \cite{Ramsay_damp:2010,Ramsay_renorm:2010} and limit the effectiveness of exciton inversion schemes using adiabatic rapid passage (ARP) at elevated temperatures \cite{Mathew:2014,Wei:2014}.  The finite size of the QD provides an opportunity to mitigate the impact of the electron-phonon interaction on the quality of optical control by limiting the range of phonon frequencies that can contribute to dephasing \cite{LukerReview}.  This limited range of coupled phonon modes, which becomes smaller the larger the QD, may be combined with the use of short laser pulses to suppress resonant phonon transitions by increasing the energy separation between the dressed states beyond the coupled phonon frequency band \cite{Mathew:2014,Vagov:2007,Luker:2017,Glassl:2011,Kaldewey:2017}.  

Here we exploit this feature to demonstrate full suppression of phonon-mediated decoherence in large QDs emitting at 1.3~$\mu$m.  The use of large, telecom-compatible QDs in our experiments is not only attractive for future long-distance quantum communication in distributed quantum networks \cite{Atature:2018,Senellart:2017}, but also dramatically reduces the strength of the laser field needed to reach the decoupling regime for the electron-LA phonon interaction.  Through calculations of the quantum state dynamics, we analyze the dependence of the threshold pulse area for decoherence suppression ($\Theta_{T}$) on the size and shape of the QD and contrast $\Theta_{T}$ for resonant and quasi-resonant control schemes.  While the latter schemes offer considerably larger brightness in single-photon source applications \cite{Senellart:2017}, we show that larger pulse areas are required to reach the decoupling regime due to the differing wave function symmetry between the QD ground and first excited state. The demonstration of low-threshold, high-fidelity inversion in our experiments points to the potential for practical QD-based quantum light sources that could be initialized in parallel and operated at elevated temperatures. 

We demonstrate decoherence suppression using ARP, for which frequency-swept optical control pulses are used to invert the exciton transition in the QD \cite{Loy:1974}.  The control pulse is given by $E(t)=\frac{1}{2}E_p(t)\exp{[-i(\omega_l t+ \alpha t^2)]}$, where $E_p(t) =  E_0 \exp(-2\ln(2)t^2/\tau_0^2)$ is the pulse envelope, $\omega_l$ is the center frequency, and $\alpha$ is the temporal chirp.  The energies ($E_{\pm}$) of the dressed states of the QD in the presence of this control pulse ($\ket{\Psi_\pm}$) are depicted in Fig.~\ref{fig:Figure1}(a).  The splitting between the dressed states is given by $\sqrt{\Omega(t)^2 + \Delta(t)^2}$, where $\Omega(t)=\frac{\mu E_p(t)}{\hbar}$ and $\Delta(t)=-2\alpha t$ are the instantaneous values of the Rabi frequency and the detuning of the laser field from the exciton transition, respectively.  For a sufficiently strong pulse intensity and chirp, the control process is adiabatic and the system remains in one of the two dressed states while the admixture of the bare QD states ($\ket{0}$ and $\ket{1}$) evolves resulting in inversion of the two-level system.  The sign of $\alpha$ determines which of the two dressed states $\ket{\Psi_\pm}$ the system traverses as it evolves from $\ket{0}$ to $\ket{1}$.

\begin{figure}[!t]
	\includegraphics[width=0.882\columnwidth]{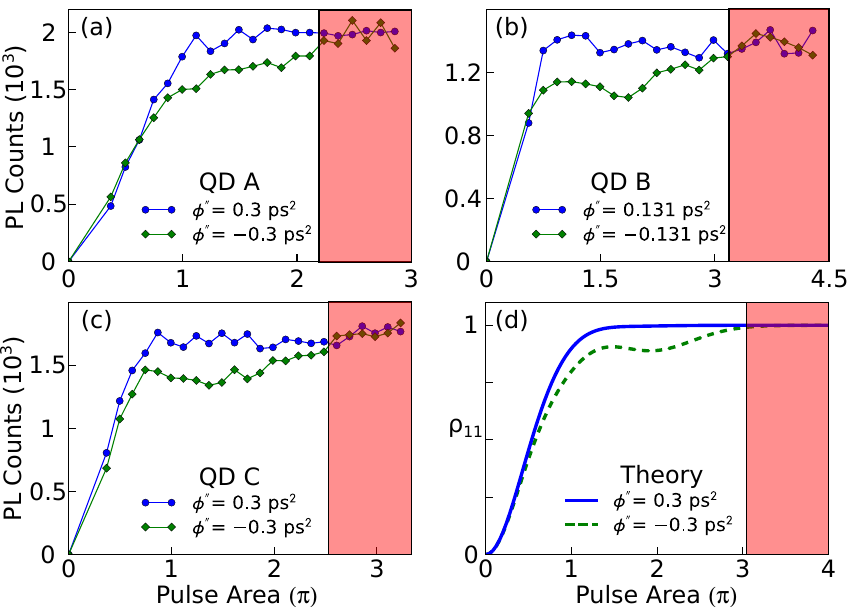}
\centering
    \caption{(a)-(c) PL intensity versus pulse area for positively-chirped (blue circles) and negatively-chirped (green diamonds) control pulses for three different QDs with GS emission wavelengths at 1259 nm, 1273 nm and 1265 nm, respectively. (d) Calculated exciton occupation for chirped-pulse excitation on the ES transition for a QD with d$_\perp$= 6.0 nm, d$_\parallel$ = 7.3 nm. Shaded area indicates the range of pulse areas with suppressed exciton-LA phonon coupling.
}
    \label{fig:Figure2}
\end{figure}

LA-phonons can cause resonant transitions between the dressed states during the optical pulse (depicted by dashed arrows in Fig.~\ref{fig:Figure1}(a)), representing the dominant contribution to dephasing \cite{Krummheuer:2002,Vagov:2002}.  These transitions can occur provided that phonons with energies equal to the dressed state splitting are coupled to the exciton \cite{LukerReview}.  The exciton-LA phonon coupling strength is dictated by the phonon spectral density $J(\omega) = \sum_\mathbf{q}  |g_\mathbf{q}|^2  \delta (\omega -\omega_\mathbf{q})$ where $\omega_{\mathbf{q}}$ = $c_sq$  is the LA phonon frequency for bulk GaAs phonon modes \cite{Stock:2011}.  The coupling strength is given by $g_\mathbf{q}  = \frac{q}{\sqrt{2V\rho \hbar \omega_\mathbf{q}}}(D_e - D_h)P[\psi(\mathbf{r})]$, where $D_{e(h)}$ is the deformation potential constant for the electron (hole), $V$ is the volume of the unit cell, $\rho$ is the mass density, $c_s$ is the speed of sound, and $P[\psi(\mathbf{r})] =  \int d^3\mathbf{r} | \psi (\mathbf{r}) |^2 e^{i\mathbf{q}\cdot \mathbf{r}} $ is the form factor for the QD.  For the simplest case of GS pumping in a spherical QD, $J(\omega) = A \omega^3 e^{-(\omega / \omega_c )^2}$ [Fig.~\ref{fig:Figure1}(b), inset], where  $\omega_c = 2\sqrt{2}c_s/d$ is a cutoff frequency that depends on $d$, the diameter of the QD.  The smaller the QD, the larger $J(\omega)$ and $\omega_c$ (i.e. the stronger the coupling at a given phonon frequency and the wider the bandwidth of coupled phonon modes).    

At any instant during the control pulse for which the dressed state splitting is much larger than $\omega_c$, the electron and LA-phonons are decoupled and dephasing is suppressed.   This decoupling regime has been studied theoretically for both Rabi rotations \cite{Vagov:2007,Glassl:2011} and ARP \cite{Luker:2017}.  For ARP, the use of a frequency-swept optical pulse leads to a nonzero dressed state splitting at all times during the control process.  This has the consequence that complete suppression of decoherence is possible for large enough values of $\Omega(t)$ and $\Delta(t)$. This suppression occurs at smaller pulse areas when large bandwidth pulses are used.  Importantly, this suppression occurs at all temperatures \cite{Kaldewey:2017}, with substantial implications for the development of practical QD-based quantum emitters for distributed quantum networks.  For such applications, a low threshold pulse intensity for decoherence suppression is desirable to lower the power requirements of the quantum network and to facilitate parallel quantum state initialization of quantum emitters using a single laser source \cite{Giesz:2015,Reindl:2017,Delteil:2016,Stockhill:2017,Thoma:2017}.  Due to the dependence of the phonon spectral density on the form factor, one can engineer the rate of decoherence tied to exciton-LA phonons and the threshold pulse area required to enter the decoupling regime by varying the size and shape of the QD.

\begin{figure}[!t]
	\includegraphics[width=0.802\columnwidth]{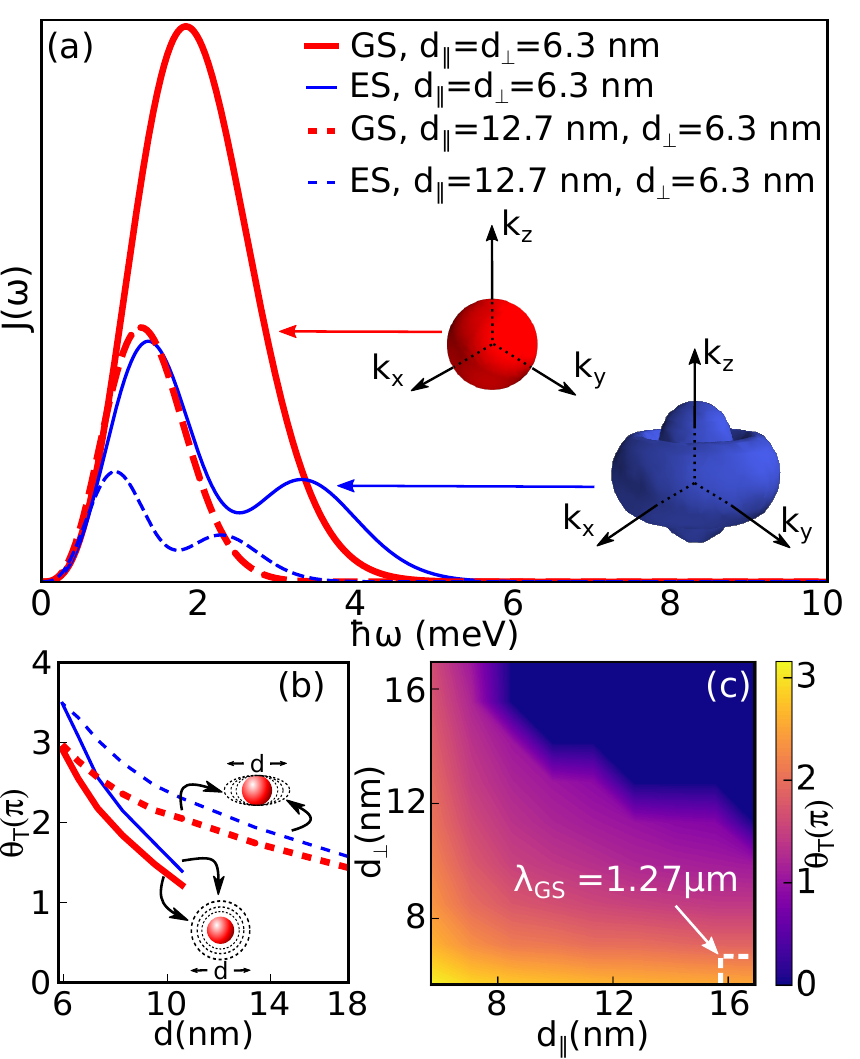}
\centering
    \caption{(a) Comparison of the calculated phonon spectral density $J(\omega)$ for optical driving of excitons on the GS and ES optical transitions.  Thick (thin) curves show $J(\omega)$ for the GS (ES), with solid (dashed) curves corresponding to results for a spherical (lens-shaped) QD.  Inset: Electron-hole form factors for GS and ES excitons in a spherical QD. (b) Calculated threshold pulse area for GS and ES excitons in QDs of varying size and shape. (c) Calculated threshold pulse area for GS excitons, with d$_\parallel$ (d$_\perp$) corresponding to the QD in-plane diameter (height).  The dimensions and threshold pulse areas corresponding to the QDs used in our experiments are indicated by the white dashed square.
}
    \label{fig:Figure3}
\end{figure}

The decoupling regime for exciton-LA phonon coupling may be detected experimentally by comparing the exciton inversion for positive and negative values of pulse chirp \cite{Mathew:2014}.  As shown in Fig.~\ref{fig:Figure1}(a), for negative (positive) chirp, the system evolves from $\ket{0}$ to $\ket{1}$ on the higher- (lower-) energy dressed state. At low temperatures, phonon emission contributes to dephasing by inducing downward transitions between the dressed states but phonon absorption is suppressed. Phonon coupling therefore impacts the quantum control process only for state evolution on the higher-energy dressed state (i.e. for negative pulse chirp).  Fig.~\ref{fig:Figure1}(b) shows the calculated chirp-sign dependence of the exciton occupation.   The difference between the exciton occupations for positive and negative pulse chirp varies with pulse area with a maximum when the dressed state splitting matches the peak value of $J(\omega)$.  For pulse areas well above $\omega_c$, the optically-driven system is decoupled from LA-phonons. To quantify the onset of the decoupled regime in the theoretical calculations, we define the threshold pulse area $\Theta_T$ as that for which the exciton occupation for negative chirp is within 1\% of that for positive chirp. 

Adiabatic rapid passage was carried out on an In(Ga)As/GaAs QD structure grown using molecular beam epitaxy under conditions to generate large QDs with a ground state emission near 1.3 $\mu$m at cryogenic temperatures \cite{Gamouras:2013}.  The ensemble photoluminescence (PL) is shown in Fig.~\ref{fig:Figure1}(c).  The emission peaks from individual QDs detected using microPL experiments are shown in Fig.~\ref{fig:Figure1}(d).  A schematic diagram of the experimental setup used for quantum control experiments is shown in Fig.~\ref{fig:Figure1}(e).   The first excited state (ES) optical transition in each QD was resonantly excited and the PL emitted from the ground-state (GS) optical transition was detected following nonradiative relaxation.  The time scale for this relaxation process (15 ps \cite{Gundogdu:2005}) is longer than the 7 ps chirped control pulse and so does not contribute significantly to exciton dephasing during optical control. This nonresonant pumping configuration enables spectral rejection of the scattered pump light. Further details on the experiments are in the Supplemental Materials.    

The results of ARP experiments on three different QDs are shown in Fig.~\ref{fig:Figure2}(a)-(c).    The convergence of the PL intensities for negative and positive pulse chirp for all QDs indicates complete suppression of decoherence tied to electron-LA phonon coupling. The threshold pulse area required to reach the decoupling regime varies between QDs due to their slight differences in size and/or shape.  These differences are also reflected in the variation in the GS emission wavelength (1.259 $\mu$m to 1.273 $\mu$m for the QDs in Fig.~\ref{fig:Figure2}). The calculated chirp sign dependent state dynamics for a lens-shaped QD (d$_\parallel$ = 7.3~nm, d$_\perp$ = 6 nm) are shown in Fig.~\ref{fig:Figure2}(d), capturing the general trends seen in the experimental curves.  The measured values of $\Theta_{T}$ are in the range 2.2$\pi$ to 3.5$\pi$.  

Decoherence suppression is observed in our experiments at much lower pulse areas than in previous quantum control experiments on QDs \cite{Mathew:2014,Ramsay_damp:2010,Ramsay_renorm:2010,Kaldewey:2017,Kaldewey_biex:2017}.  In experimental demonstrations of Rabi rotations, which have reached pulse areas as large as 14$\pi$ \cite{Ramsay_damp:2010,Ramsay_renorm:2010}, the maximum Rabi frequencies used were large enough to detect the nonmonotonic dependence of damping on pulse area near the peak of $J(\omega)$ but were still well below the decoupling regime due to the finite turn-on time of the pulse \cite{Ramsay_damp:2010,Ramsay_renorm:2010}.  In recent experiments involving ARP on excitons \cite{Kaldewey:2017}, the decoupling regime was reached at the end of the range of experimentally-accessible pulse areas, corresponding to a value of $\Theta_T$ exceeding 8$\pi$.  The 4-fold smaller values of $\Theta_T$ in Fig.~\ref{fig:Figure2} reflect the much larger QDs used in our experiments, with GS emission wavelengths close to 1.3~$\mu$m in contrast to the 930-950~nm emitting QDs used in previous work \cite{Ramsay_damp:2010,Ramsay_renorm:2010,Kaldewey:2017,Kaldewey_biex:2017}.  

Our quantum control experiments were carried out under quasi-resonant pumping (i.e. the laser was tuned to the ES transition in each QD).  Both resonant and quasi-resonant excitation schemes have been used in high-performance single photon sources \cite{Giesz:2015,Reindl:2017}.   As both pumping schemes suffer from decoherence tied to LA-phonon coupling and previous theoretical models have only considered the case of GS pumping \cite{LukerReview}, it is instructive to compare the two pumping schemes with regard to the coupling strength and threshold pulse area required for decoherence suppression.  Fig.~\ref{fig:Figure3}(a) shows the calculated spectral density for GS and ES wave functions in both spherical and lens-shaped QDs.  For these calculations, the ES transition corresponds to in-plane orbital angular momentum states.  (For more information, see the Supplemental Material.) The size dependence of $J(\omega)$ for the GS exciton is consistent with earlier studies showing stronger coupling in the spherical QD due to the larger degree of quantum confinement \cite{Luker:2017}.  There are two notable differences between the spectral density for optical driving of the GS and the ES: $J(\omega)$ for the ES exhibits a double peak structure and is characterized by a larger cutoff frequency.  These differences may be traced back to the form factors for the GS and ES wave functions [Fig.~\ref{fig:Figure3}(a), inset].   The larger bandwidth of $J(\omega)$ for ES pumping leads to a larger threshold pulse area for decoherence suppression.  Fig.~\ref{fig:Figure3}(b) and Fig.~\ref{fig:Figure3}(c) show the results of calculations of $\Theta_T$ for QDs with various QD shapes and sizes.  For a given choice of QD dimensions, $\Theta_T$ is larger for ES pumping than for GS pumping.   These results imply that resonant excitation schemes would offer an advantage over near resonant pumping geometries of lower power requirements for quantum network applications.  Our findings also indicate that the much lower values of $\Theta_T$ observed in optical control experiments involving large QDs in this work in comparison to previous experiments \cite{Ramsay_damp:2010,Ramsay_renorm:2010,Kaldewey:2017,Kaldewey_biex:2017} are expected to be even lower if GS pumping is used.  

An ideal quantum emitter will be designed to minimize $\Theta_T$ while maintaining a GS transition that is compatible with standard telecommunications wavelengths.  The results in Fig.~3(c) indicate that both the height and in-plane dimensions of the QD may be used to engineer $\Theta_T$.  The dimensions corresponding to the QDs used in this work, characterized by $\Theta_T$ between 2$\pi$ and 3$\pi$ [white box in Fig.~\ref{fig:Figure3}(c)], represent a near-optimum case for the InGaAs/GaAs family of QDs emitting near 1.3~$\mu$m.  Lower values of $\Theta_T$ would be expected for QDs in the InAsP/InP family \cite{Haffouz:2018} with weaker quantum confinement and compatibility with the low-loss telecommunication band at 1.55$\mu$m. 

In summary, we demonstrate complete suppression of decoherence for optically-mediated exciton inversion using ARP on large QDs with telecom-compatible emission wavelengths.  By exploiting the ability to engineer the strength of electron-phonon coupling by varying the size and shape of the QD, we achieve a four-fold reduction in the threshold pulse area required to reach the decoupling regime.  Our observation of low-threshold decoherence suppression would facilitate parallel quantum state inversion in single and entangled photon sources with the potential for operation at elevated temperatures\cite{Giesz:2015,Reindl:2017,Delteil:2016,Stockhill:2017,Thoma:2017}, while the telecom-compatible emission wavelengths of our QDs would support the development of distributed quantum networks incorporating long-distance quantum state transfer \cite{Atature:2018}.  Our calculations indicate a new direction for engineering the strength and bandwidth of coupling via the symmetry of the QD wave function.  These findings may aid efforts to exploit coherent phonon wave packet emission for coupling the quantum states of neighbouring QDs \cite{Wigger:2014,Luker:2017}. Our calculations also indicate that even lower threshold pulse areas for decoherence suppression than those reported here may be expected for resonant pumping schemes. Our results will support the application of semiconductor QDs in future distributed quantum networks.  

This research is supported by the Natural Sciences and Engineering Research Council
of Canada.

\end{document}